\documentclass[published]{JHEP3}

\JHEP{01(2003)070}
\received{October 13, 2002}
\revised{January 28, 2003}
\accepted{January 30, 2003}
\keywords{M(atrix) Theories, Penrose limit and pp-wave background, Supersymmetry Breaking}

\usepackage{epsfig}

\skip\footins = 1\bigskipamount plus 2pt minus 4pt

\newcommand{\U}{\mathop{\rm {}U}\nolimits}
\newcommand{\SU}{\mathop{\rm SU}\nolimits}
\newcommand{\Tr}{\mathop{\rm Tr}\nolimits}
\renewcommand{\Re}{\mathop{\rm Re}\nolimits}
\renewcommand{\Im}{\mathop{\rm Im}\nolimits}

\title{A quantum mechanical model of spherical supermembranes}

\author{John Conley, Ben Geller, Mark G. Jackson,
Laura Pomerance and Sharad Shrivastava\\
Department of Physics, Columbia University\\
New York, NY 10027, USA\\
E-mail: \email{JConley@trevornet.org}, \email{bdavidgeller@yahoo.com}, 
\email{markj@phys.columbia.edu}, \email{laurapomerance@hotmail.com}, 
\email{sharad@phys.columbia.edu}}

\abstract{We present a quantum mechanical model of spherical
supermembranes.  Using superfields to represent the cartesian
coordinates of the membrane, we are able to exactly determine its
supersymmetric vacua.  We find there are two classical vacua, one
corresponding to an extended membrane and one corresponding to a
point-like membrane.  For the ${\mathcal N} = 2$ case, instanton
effects then lift these vacua to massive states.  For the ${\mathcal
N} = 4$ case, there is no instanton tunneling, and the vacua remain
massless.  Similarities to spherical supermembranes as giant gravitons
and in Matrix theory on pp-waves is discussed.}

\begin{document}
\section{Introduction}

Spherical supermembranes are a topic which have received much
attention lately.  First quantized in light-front gauge by de Wit,
Hoppe and Nicolai~\cite{whn}, they have become relevant in the study
of holography on $AdS_n \times S^m$
spaces~\cite{ads,mcgreevy,goliath}.  Recent studies of Matrix theory
on pp-wave backgrounds has also resulted in studying
supermembranes~\cite{pp,sup1,sup2,nori}, although the results are
somewhat obfuscated by the large-component spinors arising from
dimensional reduction.  Thus a simple model, capturing the essentials
of the supermembrane dynamics, would be useful.

In this paper we present a simplified model of spherical
supermembranes, based on the quantum mechanics of 3 superfields and 2
or 4 supercharges.  In the classical approximation, where
quantum-mechanical instanton tunneling effects are negligible, there
are 2 supersymmetric vacua present.  In the $\mathcal N = 2$ case, the
vacua have opposite statistics, allowing instanton tunneling.  We then
identify the instanton responsible for lifting these vacua to massive
states.  Thus we are left with no supersymmetric vacua.  In the
$\mathcal N = 4$ case, the vacua have similar statistics, preventing
instanton tunneling.

We finally discuss the significance of these vacua, and their possible
relation to Matrix theory compactification on pp-waves.

\section{Matrix theory membranes}

Our starting point is the result obtained by de Wit et al~\cite{whn}
and Kabat and Taylor~\cite{kabat_sph} that a $N \rightarrow \infty$
matrix model could be used to study M2-branes of spherical topology,
or ``spherical supermembranes".  The membrane action is\footnote{The
spacetime metric used initially is $\eta_{\mu \nu} = (-+ \cdots + )$.
Light-front coordinates are $X^\pm \equiv \frac{1}{\sqrt{2}} (X^0 \pm
X^{10})$.  $X^+$ is light-front time; the conjugate energy is $p^- =
\frac{1}{\sqrt{2}} (E - p^{10})$.  The coordinate $X^-$ is
compactified, $X^- \simeq X^- + 2 \pi R$, and the conjugate momentum
$p^+ = \frac{1}{\sqrt{2}} (E + p^{10})$ is quantized, $p^+ = N/R$.
The tension of the membrane is $T_2 ={1}/{(2 \pi)^2 l_P ^3}$.}
\[
S = -\frac{T_2}{2} \int d ^3 \zeta \sqrt{-g} \left( g^{\alpha \beta}
\partial _\alpha X^\mu \partial _\beta X^\nu \eta_{\mu \nu} -1 \right) 
\]
where the $\zeta^\alpha = (t, \sigma^a)$ are coordinates on the
membrane worldvolume, $g _{\alpha \beta}$ is an auxiliary worldvolume
metric, and $X^\mu (\zeta)$ are embedding coordinates.  The
light-front gauge-fixing of this action is performed
in~\cite{whn,bst}, making use of $N \times N$ hermitean matrices to
approximate the Lie algebra of area-preserving diffeomorphisms on the
sphere.  Specifically, consider the three cartesian coordinate
functions $x_1,x_2,x_3$ on the unit sphere.  The Poisson brackets of
these functions are given by
\[
\{x_A, x_B \} = \epsilon_{ABC} x_C\,. 
\]
From this we can produce the equations of motion for the transverse
coordinates $X^i$:
\[
{\ddot X}^i = \frac{4}{N^2} \{ \{ X^i, X^j \}, X^j \} 
\]
which follow from the hamiltonian
\[
H = \frac{N}{2 \pi R} \int d ^2 \sigma \left( \frac{1}{2} {\dot X}^i
{\dot X}^i + \frac{1}{N^2} \{ X^i,X^j \} \{ X^i, X^j \} \right) .
\]
We can then make a correspondence between these coordinate functions on
$S^2$ with the generators of $\SU(2)$:
\[
x_A \leftrightarrow \frac{2}{N} J_A 
\]
where $J_1, J_2, J_3$ are generators of the $N$-dimensional representation
of $\SU(2)$ satisfying
\[
-i [J_A,J_B] = \epsilon_{ABC} J_C \,.
\]
Then completing the matrix correspondence with
\[
\{ \cdot ,\cdot  \} \rightarrow \frac{-iN}{2} [\cdot ,\cdot ], \qquad  \frac{N}{4 
\pi} \int d^2 \sigma
\rightarrow \Tr  \,,
\]
we produce the hamiltonian of Matrix theory,
\begin{equation}
H = \frac{1}{R} \Tr  \left( \frac{1}{2} {\dot {\bf X}}^i {\dot
{\bf X}}^i - \frac{1}{4} \left[ {\bf X}^i, {\bf X}^j \right] \left[
{\bf X}^i, {\bf X}^j \right] \right)
\end{equation}
where ${\bf X}^i(t) = \phi_i (t) J_i $, $i=1,2,3$ as the
(noncommutative) $i$-coordinate of a supermembrane in light-cone
gauge.  The equations of motion are then:
\begin{eqnarray*}
{\ddot \phi_1} &=& - \phi_1 (\phi_2 ^2+ \phi_3^2)\,, \\
{\ddot \phi_2} &=& - \phi_2 (\phi_1 ^2+ \phi_3^2)\,, \\
{\ddot \phi_3} &=& - \phi_3 (\phi_1 ^2+ \phi_2^2)\,,
\end{eqnarray*}
It was found in~\cite{kabat_sph} that the membranes are unstable, and
the only vacuum configuration is $\phi_1 = \phi_2 = \phi_3 = 0$.  This
has a clear interpretation that the tension of the brane is forcing it
to contract to zero size.

In subsequent work, Myers~\cite{myers} found there was a term coupling
the membrane to the four-form field strength found in supergravity,
\begin{equation}
V_1 = \frac{i}{6R} F_{0ijk} \Tr  {\bf X}^i {\bf X}^j {\bf X}^k \,.
\end{equation}
As might be hoped, this can stabilize the membrane by stretching it,
the membrane being charged under this gauge field.  However, there was
no analysis of the supersymmetry of the membrane in this background
field.

At this point we should clarify what we mean by $\phi_i$ as the
cartesian coordinates of the membranes.  We define $\phi_i$ such that
a \emph{membrane} will have radii $\phi_i >0$.  As $\phi_i \rightarrow
0$ for one coordinate, we see the membrane is compressed along that
dimension until it resembles a disk.  We can then continue to $\phi_i
< 0$ if we interpret it as an \emph{antimembrane} with radius
$-\phi_i$; the intuition behind this is that the membrane has now
turned inside-out, and so we would expect to have opposite charge to
all gauge fields (its internal orientation is now reversed).  A
similar situation exists if several $\phi_i <0$, so that a membrane
can be defined as $\phi_1 \phi_2 \phi_3 > 0$ and an antimembrane has
$\phi_1 \phi_2 \phi_3 < 0$.  In fact, many solutions which appear
distinct are actually equivalent because the ${\bf X}^i$ are defined
only up to a unitary transformation,
\[
{\bf X}^i \rightarrow U^\dagger {\bf X}^i U\,. 
\]
For example, the two solutions
\begin{equation}
\label{2sols}
\phi_1,\phi_2,\phi_3 \qquad  \hbox{and} \qquad  -\phi_1,-\phi_2,\phi_3 
\end{equation}
are related by $U = \exp i \pi J_3$.

\section{Constructing the toy $\mathcal N = 2$ superpotential}

The simplest model of the supermembrane would constitute 3 real
bosonic fields $\phi_i$ representing the cartesian coordinates,
depending only on time, and their fermionic partners $\psi_i$.  Thus
this is a quantum mechanical system of 3 real superfields $\Phi_i$,
\[
\Phi_i = \phi_i + i \theta \psi_i - i \psi^*_i \theta^* + \theta \theta 
^* D_i 
\]
where $\theta, \theta^*$ are Grassmann variables and $D_i (t)$ is an 
auxiliary function.  In order to produce a superpotential capable of 
producing $V_1$, we found it was necessary for a second potential to be 
added, equivalent to a ${\bf X}^i$ mass term:
\[
V_2 = \frac{1}{2R} m^2 \sum _i {\bf X}^i {\bf X}^i\,.
\]
As explained in~\cite{kabat_lin}, this matrix theory potential term is 
physically realized by modifying the background metric $g_{++}$ component,
\[
g_{++} = - m^2 \sum_i {\bf X}^i {\bf X}^i \,. 
\]
By choosing 
\[
F_{0ijk} = 3m \epsilon_{ijk} 
\]
the entire system can be recast in the superfield formalism,
\begin{eqnarray}\label{H}
S &=& \frac{1}{R} \int d^2 \theta dt \left( \sum _i | D_\theta \Phi_i |^2 - W 
\right) \\
&=& \frac{1}{R} \int dt \sum _i\left\{ \frac{1}{2} \dot{\phi}_{i} \dot{\phi}_{i}
+i {\psi ^\dagger_i} \dot{\psi}_{i}
-\frac{1}{2} \left| \frac{\partial W}{\partial\phi_{i}}
\right|^2+ \sum_j \frac{\partial^{2} W}{\partial
\phi_{i}\partial\phi_{j}} {\psi ^\dagger_i} \psi_{j} \right\},
\\
\label{superpot}
W &=& \Phi_1 \Phi_2 \Phi_3 - \frac{m}{2} \sum _i \Phi_i ^2\,. 
\end{eqnarray}

This superpotential is similar to that studied in $\mathcal N = 1^*$
theories~\cite{n1}.  In what follows we derive equations of motion
from this superpotential and show that they are the same equations of
motion found for the Matrix hamiltonian in the previous section. After
identifying the classical vacua of this potential, we then calculate
the Witten index, suggesting the breaking of supersymmetry.  Finally,
we describe instanton tunneling between the bosonic and fermionic
vacua, confirming that supersymmetry is broken.

We first derive the potential,
\begin{eqnarray}
\label{riccati}
V(\phi_i) &=& \frac{1}{2} \sum_{i=1}^{3}\left|\frac{\partial W}{\partial
\phi_{i}}\right|^{2} 
\\
&=&\frac{1}{2} \sum_{i\neq j}\phi_i^2 \phi_{j}^{2} +\frac{m^2}{2} \sum_{i}\phi_{i}^{2} -
3 m \phi_1\phi_2\phi_{3}\,.
\end{eqnarray}
The equations of motion are then
\begin{eqnarray}
\ddot{\phi_{1}}&=& -\phi_{1}\left( \phi_{2}^{2}+\phi_{3}^{2}\right) -
m^2 \phi_{1} + 3m \phi_{2}\phi_{3}\,,
\nonumber\\
\ddot{\phi_{2}}&=& -\phi_{2}\left( \phi_{3}^{2}+\phi_{1}^{2}\right) -
m ^2 \phi_{2} + 3m \phi_{3}\phi_{1} \,,
\nonumber\\
\ddot{\phi_{3}}&=& -\phi_{3}\left( \phi_{1}^{2}+\phi_{2}^{2}\right) -
m^2 \phi_{3} + 3m \phi_{1}\phi_{2} \,.
\label{eom2}
\end{eqnarray}
These are precisely the same as the Matrix equations of motion
obtained in the previous section for a spherical membrane in
background fields $g_{\mu \nu}$ and $F_{0ijk}$, producing potentials
$V_1$ and $V_2$.  And because we obtained this from a superpotential,
we may easily determine the supersymmetric properties of the
vacuum~\cite{cooper}.

\subsection{Calculation of the Witten index}

We will first locate the classical vacua satisfying $V = 0$,
neglecting any quantum-mechanical tunneling which might raise the
vacuum energy.  From this we can determine the Witten index $\Delta$,
\[
\Delta \equiv N_+ - N_- 
\]
where $N_+$ is the number of bosonic vacua and $N_-$ is the number of
fermionic vacua.  A remarkable property of the Witten index is that it
does not change under deformation of parameters, so we expect this
will be the index value even when tunneling effects are included.

\pagebreak[3]

The superpotential has the following two distinct classical vacua,
whose labeling will become apparent shortly:
\begin{equation}\label{vacua}
\begin{array}{cccccc}
& & \phi_1 & \phi_2 & \phi_3 \\
| F \rangle \equiv& | &0,&  0, & 0 & \rangle\,, \\
| B \rangle \equiv& | & m, &  m, & m & \rangle \,.
\end{array}
\end{equation}
In listing these vacua we have neglected all physically identical
vacua which may be obtained via unitary transformations, as explained
previously.  Note that for $|B \rangle$, $\phi_1 \phi_2 \phi_3 = m^3$,
and so this is a membrane (as opposed to an antimembrane).  Physically
this arises because a certain $F_{0ijk}$ charge is required to balance
the tension and metric potentials possessing membrane-antimembrane
symmetry.  By definition, membranes have this charge, antimembranes
having the opposite charge, so we require a membrane.

To determine the type of vacua (bosonic or fermionic) each of these
are, we investigate the fermion mass term of the supersymmetric
action,
\begin{equation}\label{fermass}
\frac{\partial^2W}{\partial\phi_i\partial\phi_j}=
\pmatrix{
-m & \phi_3 & \phi_{2}\cr
\phi_{3} & -m & \phi_{1}\cr
\phi_{2} & \phi_{1} & -m }.
\end{equation}
When the matrix is evaluated for a particular vacuum, its eigenvalues
indicate the energy cost of the presence of a fermion in that ground
state.  A negative eigenvalue implies that the ground state contains a
fermion. If an odd number of eigenvalues are negative, the ground
state is fermionic. For example, the vacuum at the origin $| F
\rangle$ is fermionic:
\begin{equation}
\lambda = \{-m,\ -m,\ -m \}\,,
\end{equation}
and the extended solution $|B\rangle$ is bosonic:
\begin{equation}
\lambda = \{ m, \ -2m,\  -2m \}\,.
\end{equation}
The Witten index is therefore
\[
\Delta = 1-1 =0\,. 
\]
A non-zero Witten index implies that SUSY may or may not be broken.
Although we obtained this answer by studying the vacua individually,
in the $m \rightarrow \infty$ limit, the index should not change as we
reduce $m$ and ``turn on" quantum mechanical interactions.

\subsection{Instanton tunneling}

The value of the Witten index can be confirmed and elucidated by
considering the instanton tunneling that lifts certain linear
combinations of these vacua. Instanton effects in supersymmetric
quantum mechanics have been well-studied~\cite{inst}.

We seek the instanton solution with the following properties.  At
euclidean time $t \rightarrow -\infty$, it begins at the vacuum $| F
\rangle = | 0,0,0 \rangle$.  As time progresses it moves towards the
vacuum $| B \rangle = | m,m,m \rangle$, arriving there as $t
\rightarrow \infty$.  Since the solution should be symmetric in all
$\phi_i$, we may use the following action to derive the instanton
solution:
\begin{equation}
\label{eucaction}
S_E = \frac{3}{2} \int dt \left({\dot \phi}^2 + V(\phi) + \cdots \right)
\end{equation}
where $\cdots$ indicate the fermion terms, which play no role in our
calculation, and the potential is given by
\[
V(\phi) = \phi^2 (\phi - m)^2\,. 
\]
The instanton solution then satisfies
\[
{\dot \phi}_{I} = -\phi_{I} (\phi_{I} - m)\,. 
\]
Choosing $t_I$ as the time at which the instanton is half-way 
between the two vacua, this is easily solved to yield
\begin{equation}
\label{inst}
\phi _{I} (t)= \frac{m}{1 + e^{-m(t-t_I)}}\,.
\end{equation}
Plugging this back into~(\ref{eucaction}), we produce the classical
action
\[
A_0 = \frac{m^3}{2}\,. 
\]
Hence these classical vacua each acquire a mass $\sim e^{-m^3/\hbar}$.

\section{Constructing the toy $\mathcal N = 4$ superpotential}

We now extend the number of supercharges to 4 simply by complexifying
the superfields $\Phi_i$ and parameter $m$.  In particular, the
superpotential~(\ref{superpot}) and its vacua are identical.  And
again, unitary transformations may be used to bring all extended
solutions to the form $| m,m,m\rangle$ where the basis $|
\phi_1,\phi_2,\phi_3 \rangle$ is now complex.  The membrane will then
have radius $|m|$, with the phase of $m$ determining the complex
basis.

For the special case $m=0$, this model reproduces the `rotating
ellipsoidal membrane' analysis in~\cite{harsav}, where the membrane
was stabilized by adding angular momentum to the complex fields.  For
$m \neq 0$, our potential does not have this $\U(1) \times \U(1)
\times \U(1)$ symmetry.

We may choose the convention that 
the six mass eigenvalues for the state $| m,m,m\rangle$,
\[
\lambda = \{ \Re m,\Im  m,-2 \Re  m,-2 \Im  m, -2\Re 
m,-2\Im  m \}
\]
constitute a bosonic state.   The point-size solution $|0,0,0 \rangle$ now has 
mass eigenvalues 
\[
\lambda = \{-\Re  m,-\Im  m,-\Re  m,-\Im  m,-\Re  m,-\Im  m\}\,,
\] 
which makes it also a bosonic state.  Since there cannot be tunneling 
between vacua of like statistics, these vacua remain massless.

\section{Discussion}

We have presented a quantum mechanical model of the spherical
supermembrane.  For either 2 or 4 supercharges, it possesses two
supersymmetric vacua, but in the former case the vacua acquire mass
when instanton effects are included.

This situation is similar to that believed to occur in the study of
giant gravitons~\cite{mcgreevy,goliath}.  In fact, the instanton
solution~(\ref{inst}) we obtained using quantum mechanics is identical
to the exact field theory result obtained in~\cite{goliath}.  Perhaps
our model could be modified to include a second extended membrane
vacua, as is the case in giant gravitons.

The most interesting aspect of the model is that the background fields
required for our toy model bear a striking resemblence to Matrix
theory compactified on a pp-wave background with our parameter $m$
equivalent to the parameter $\mu/3$ found in~\cite{pp} (in particular,
the metric and field strengths are identical).  Ours differs only in
that we limit ourselves to the 3 relevant bosonic fields (playing the
role of the cartesian coordinates), our fermions are Grassmann numbers
and carry no spinor indices arising from dimensional reduction, and
our supercharges commute with the hamiltonian.  It would be very
interesting to determine whether our model can be obtained as a
special case of supersymmetric Matrix theory compactification.

\acknowledgments

We would like to thank D.~Berenstein, R.~Friedman, B.~Greene,
N.~Iizuka, K.~Schalm, A.~Weltman and especially D.~Kabat. This work
was supported by the VIGRE program.  M.G.J. is supported by a Pfister
Fellowship.

\end{document}